\documentclass{article}
\usepackage{spconf,amsmath,graphicx}

\usepackage{enumitem}
\setlist{nosep, leftmargin=14pt}

\newcommand{\dd}{\scriptsize{\textnormal{d}}}

\usepackage[utf8]{inputenc}
\usepackage{graphicx}
\usepackage{xcolor}
\usepackage{overpic}
\usepackage{wrapfig,lipsum}

\usepackage{amsmath, amsthm, amssymb, amsfonts}
\usepackage{array}
\usepackage{nicefrac}
\usepackage{hyperref}
\usepackage{algorithm,algorithmic}

\title{A Nonlinear Hierarchical Model for Longitudinal Data on Manifolds}
\name{Martin Hanik$^{\dagger}$ \qquad Hans-Christian Hege$^{\star}$ \qquad Christoph von Tycowicz$^{\dagger}$}

\address{$^\dagger$ Freie Universit\"at Berlin, Germany \qquad $^{\star}$ Zuse Institute Berlin, Germany}
\begin{document}
%
\maketitle
\begin{abstract}
Large longitudinal studies provide information that is particularly valuable in medical studies. A problem that must be solved in order to realize the full potential is the correlation between intra-subject measurements taken at different times. For data in Euclidean space this can be achieved with hierarchical models, i.e., models that account for intra-subject and between-subject variability at two different levels. However, data from medical studies often take values in nonlinear manifolds. For such data, as a first step, geodesic hierarchical models have been developed that generalize the linear ansatz by assuming that time-induced intra-subject variations occur along a generalized straight line in the manifold. However, this assumption often does not hold, for example in periodic motions or processes with saturation. As a more general alternative we propose a hierarchical model for manifold-valued data that incorporates trends along higher-order curves, namely Bézier splines in the manifold.
To this end, we present a principled way of comparing shape trends in terms of a functional-based Riemannian metric.
Remarkably, this metric allows efficient, yet simple computations by virtue of a variational time discretization that only requires solving regression problems. 
We validate our model using longitudinal data from the Osteoarthritis Initiative, including classification of disease \emph{progression}.
\end{abstract}

\section{Introduction}
    In medicine, longitudinal studies are of great importance, as they provide information on developmental phenomena that may allow improved prognoses and thus more targeted therapies.
    To address the problem of strong correlation between measurements taken from the same individual at different times, multivariate hierarchical models were developed for data from Euclidean spaces.
    In recent years, the enormous potential of longitudinal studies has also come into focus in the analysis of shape and appearance data~\cite{GerigFishbaughSadeghi2016}.
    In order to obtain as much information as possible from such data, it is necessary to leave the realm of Euclidean vector spaces and turn to methods from Riemannian geometry, as curved manifolds are their natural domains. That is, standard methods from multivariate statistics for analyzing longitudinal data must be transferred to these more general spaces. For hierarchical models this has been done only partially so far. 
    Current parametric hierarchical models for manifold-valued data are based either on geodesics (i.e., generalized straight lines)~\cite{MuralidharanFletcher2012,Singh_ea2013,Singh_ea2016,BoneColliotDurrleman2018} or general trajectories~\cite{ChakrabortyBanerjeeVemuri2017,SrivastavaKlassen2016}; further, an approach based on nonparametric curves was presented in~\cite{CampbellFletcher2018}. As an alternative to the geodesic models, the authors of~\cite{NavayazdaniHegevonTycowicz2019} proposed to use a different Riemannian structure that allows faster algorithms for high-dimensional data.
    However, to our knowledge, there is no \emph{in-between} hierarchical model for parametric trends of higher-order (i.e., based on generalized polynomials) that (a) allows to model more complicated trends and (b) is efficient enough to handle large data bases due to a small number of degrees of freedom.
    Since many phenomena, e.g., cyclic motion of cardiac anatomy, are only inadequately characterized by geodesic models, such a model could be useful in numerous applications.
    
    Therefore, in this paper, we propose a \emph{higher order hierarchical model} for the analysis of longitudinal manifold-valued data.
    By modelling, in a first step, subject-wise trends as splines consisting of generalized Bézier curves~\cite{PopielNoakes2007} (i.e., generalized polynomials), we are able to capture a vast range of phenomena---not least periodic ones.
    To obtain the trajectories from the data, we rely on regression with Bézier splines developed in~\cite{Hanik_ea2020}. In the second step, the obtained trajectories are considered as perturbations of a common mean (curve).
    To this end, we adapt the functional-based metric from~\cite{NavayazdaniHegevonTycowicz2019} to compare the obtained subject trajectories within the \emph{space of Bézier splines}.
    Thereby, we are able to analyze the inter-subject variability without interference from correlated measurements.
    For efficient calculations in the obtained space of Bézier splines, we rely on the geodesic calculus introduced in~\cite{Heeren_ea2018,RumpfWirth2014} and also used in~\cite{NavayazdaniHegevonTycowicz2019}.
    An implementation of the presented approach is publicly available as part of the open-source library \emph{Morphomatics}~\cite{Morphomatics}.
    
    We validate our model using data from the Osteoarthritis Initiative. To the best of our knowledge, this is the first time that whole trajectories (representing progression of osteoarthritis) and not only momentary states are classified.

\section{Hierarchical model}

\subsection{Bézier curves in Riemannian geometry}
    
    For preparation we first recall some facts from Riemannian geometry and about Bézier curves on manifolds. 
    In the following ``smooth'' means ``infinitely often differentiable''.
    
    A Riemannian manifold is a differentiable manifold $M$ together with a Riemannian metric $\langle \cdot,\cdot \rangle_p$ that assigns to each tangent space $T_p M$ a smoothly varying scalar product; the metric also induces a distance function $d$.
    For every Riemannian manifold there is a unique Levi-Civita connection $\nabla$. Given two vector fields $X,Y$ on $M$, it allows us to differentiate $Y$ along $X$; the result is again a vector field, which we denote by $\nabla_X Y$.
    The connection allows us to define geodesics (i.e., generalized straight lines). A curve $\gamma$ is called geodesic if its acceleration vanishes identically, i.e., $\nabla_{\gamma'}\gamma' = 0$, where $\gamma' := \frac{\dd}{\dd t} \gamma$.
    It is a useful fact that each element of $M$ has a so-called normal convex neighbourhood $U$. Any two points $p,q \in U$ can be joined by a unique length-minimizing geodesic $[0,1] \ni t \mapsto \gamma(t;p,q)$ that does not leave $U$. Throughout this paper, we always assume to work in a convex neighbourhood in order to use this property. 
    
    Next, we recall manifold-valued Bézier curves and splines. For clarity, we restrict the domain of definition of Bézier curves to $[0,1]$ in this work; in general, the curves can always be reparametrized.
    
    A set of $k+1$ control points $p_0,\dots,p_k \in U$
    defines a continuously differentiable \textit{Bézier curve} $\beta:[0,1] \to M$ \textit{of order} $k$ according to the \textit{generalized de Casteljau algorithm}
    \begin{align*}
        \beta_i^0(t) &:= p_i, \nonumber \\
        \beta_i^l(t) &:= \gamma(t;\beta_i^{l-1}(t),\beta_{i+1}^{l-1}(t)), \\
        & \quad \quad \quad \quad \quad \quad \quad l=1,\dots,k, \quad i=0,\dots,k-l,
        \label{eq: deCasteljau}
    \end{align*}
    by $\beta(t) := \beta_0^k(t)$; see Ref.~\cite{PopielNoakes2007}.
    Several such curves, say $L$, can be joined to a continuously differentiable spline~\cite{GousenbourgerMassartAbsil2019}: For $i=0,\dots,L-1$ let $p^{(i)}_0,\dots,p^{(i)}_{k_i}$ be the control points of the curves with
    \begin{equation} \label{eq: C1_condition}
        p^{(i)}_{k_i} = p^{(i+1)}_0 \quad \text{and} \quad  \gamma \left( \frac{k_i}{k_i + k_{i+1}};p^{(i)}_{k_i-1},p^{(i+1)}_1 \right) = p^{(i+1)}_0
    \end{equation}
    for all $i$ but $0$ and $L-1$. 
    Then, the corresponding \emph{Bézier spline} $B$ is defined by
        $B(t) := \beta(t-i;p^{(i)}_0,\dots,p^{(i)}_{k_i}), \quad t \in (i,i+1]. $
    
    If $L > 1$ and the first and last segment of $B$ are at least cubic, $B$ can be closed. Then, $B$ is $C^1$ and closed if and only if (\ref{eq: C1_condition}) extends cyclically as discussed in~\cite{Hanik_ea2020}.

    In the following, we set 
        $K :=  k_0+k_1+\cdots+k_{L-2}+k_{L-1}$ if $B$ is not closed ($K :=  k_0+k_1+\cdots+k_{L-2}+k_{L-1} - 1$ if $B$ is closed)
    and denote the set of $K+1$ \textit{distinct} control points of $B$ by $p_0,\dots,p_K$. In the non-closed case this means
    \begin{align*}
        (p_0,\dots,p_K) := \Big(p^{(0)}_0, &\dots,p^{(0)}_{k_0},p^{(1)}_1,\dots,p^{(1)}_{k_1}, \dots, \\ &p^{(L-1)}_{1},\dots,p^{(L-1)}_{k_{L-1}} \Big) \in U^{K+1},
    \end{align*}
    while $p_0^{(0)}$ is left out for closed $B$.
    
    Regression with intrinsic Bézier splines~\cite{Hanik_ea2020} models the relationship between an independent scalar variable and an $M$-valued dependent variable as a Bézier spline (with a fixed number of control points). That is, for $N$ data pairs $(t_i, q_i) \in [0,1] \times U$, the minimizer (represented by its control points) of the sum-of-squared energy
    \begin{equation} \label{eq:regression_objective}
          \mathcal{E}(p_0,\dots,p_K) := \frac{1}{2} \sum^N_{j=1} d \Big(B(t_j;p_0,\dots,p_K),q_j \Big)^2
    \end{equation}
    models the relationship between $t$ and $q$.

\subsection{The Model}
    In this section we introduce the nonlinear hierarchical model.
    For this, we define the set of Bézier splines in a normal convex neighbourhood $U$ with a given number of segments and degrees:
    \begin{align*}
        \mathcal{B}^L_{k_1,\dots,k_L} := \{B: &[0,L]\to U\ |\ B \text{ is $C^1$ Bézier spline} \\ &\text{ with $L$ segments of degrees $k_1, \dots, k_L$}\}.
    \end{align*} 
    In the following we assume $L$ and $k_1,\dots,k_L$ to be fixed and, hence, omit the indices for readability.
    
    Consider that for $S$ subjects $N_s$ measurements of an independent scalar variable and a manifold-valued dependent variable are given, that is, 
    $$\left( t_i^{(s)}, q_i^{(s)} \right) \in \mathbb{R} \times U, \quad  i=1,\dots,N_s, \quad s=1,\dots,S.$$
    (Note that the number $N_s$ of measurements can be different for each subject.)
    Such data can, for example, arise in a longitudinal study that observes shape developments in several individuals.
    
    In a \emph{first step}, we model the individual trends by regression with Bézier splines of fixed type; that is, for each $s=1,\dots,S$ we perform spline regression with respect to the data $( t_i^{(s)}, q_i^{(s)})$ and, thus, obtain Bézier splines $B^{(s)} \in \mathcal{B}$, that represent the intra-subject trends.
    
    In the \emph{second step}, we model the individual trends as perturbations of a \emph{common mean trajectory}. 
    We do this by considering $\mathcal{B}$ as a submanifold of the manifold of all smooth curves in $M$. Through the identification of the curves and their control points, tangent vectors at the control points naturally translate into \emph{generalized Jacobi fields} along the curves~\cite{BergmanGousenbourger2018}. The metric from Ref.~\cite[Sec.~3.3]{SrivastavaKlassen2016} can thus be restricted to $\mathcal{B}$ to yield a natural Riemannian metric. In particular, it induces a natural distance between two Bézier splines that can be efficiently evaluated using variational time-discretization~\cite{RumpfWirth2014}, as described in the following.
    
\subsection{Computation}
    
    Let $B_1, B_2 \in \mathcal{B}$. A path between $B_1$ and $B_2$ through $\mathcal{B}$ may be represented as a parametrized surface in $M$ because it induces a map $H:[0,1] \times [0,L] \to U, (s, t) \mapsto H(s,t)$ with $H(0, \cdot) = B_1$ and $H(1,\cdot) = B_2$.
    A geodesic between $B_1$ and $B_2$ is then defined as the minimizer of the \emph{path energy}
    $E(H) := \int_0^1 \int_0^L \langle \textnormal{d} H / \textnormal{d}s, \textnormal{d} H / \textnormal{d} s \rangle \textnormal{d}t\, \textnormal{d}s.$
    Discretizing in $\mathcal{B}$ and identifying splines with their control points, we obtain a \emph{discrete $n$-geodesic} $(p_0^j,\dots,p_K^j)_{j=0,\dots,n} \in (U^{K+1})^{n+1}$ between $B_1$ and $B_2$ as the minimizer of the \emph{discrete path energy}
    $E_n((p_0^j,\dots,p_K^j)_{j=1,\dots,n}) := n\sum_{j=1}^{n-1} \int_0^L d ( B \big(t; p_0^j,\dots,p_K^j),$ $B (t; p_0^{j+1},\dots,p_K^{j+1}))^2 \textnormal{d}t.$
    The integral can be evaluated using a suitable quadrature rule. In order to approximate the minimizer of the discrete energy, we extend the iterative procedure from Ref.~\cite{NavayazdaniHegevonTycowicz2019} to our setting: We compute the discrete $n$-geodesics between two curves by iteratively performing spline regression. First, we initialize the control points of the inner curves equidistantly along the geodesics that connect the corresponding control points of $B_1$ and $B_2$. Then, the inner curves are updated so that they lie ``in the middle'' of their neighbors; to this end, we replace them with the result of spline regression with respect to $K+1$ data points that are given by (equidistant) evaluations of the neighboring curves. The procedure is summarized in Alg.~\ref{algo:discgeodesic}.
    
    \begin{algorithm}
        \hspace*{\algorithmicindent} \textbf{Input:} $B_1, B_2 \in \mathcal{B}$ with control points $\left(p_0^0,\dots,p_K^0 \right), \left(p_0^n,\dots,p_K^n \right)$, respectively\\
        \hspace*{\algorithmicindent} \textbf{Output:} Control points $(p_0^{j},\dots,p_K^{j})_{j=0,\dots,n}$ of the discrete $n$-geodesic from $B_1$ to $B_2$  
        \begin{algorithmic}[H]
            \FOR{$j=1,\dots,n-1$}
                \STATE $(p_0^j,\dots,p_K^j) \gets \left( \gamma\left( \frac{j}{n};p_0^0,p_0^n \right), \dots, \gamma\left( \frac{j}{n};p_K^0,p_K^n \right)  \right)$
            \ENDFOR
            \REPEAT
                \FOR{$j=1,\dots,n-1$}
                    \FOR{$i = 0, \dots, K$}
                        \STATE $\left( t_i^{j-1},q_i^{j-1} \right) \gets \left( \frac{iL}{K},B\big( \frac{iL}{K};p^{j-1}_0,\dots,p^{j-1}_K \big) \right)$
                        \STATE $\left( t_i^{j+1},q_i^{j+1} \right) \gets \left( \frac{iL}{K},B \big( \frac{iL}{K};p^{j+1}_0,\dots,p^{j+1}_K \big) \right)$
                    \ENDFOR
                    \STATE $(p_0^{j},\dots,p_K^{j}) \gets \texttt{reg} \Bigg( \left( t_i^{j-1},q_i^{j-1} \right)_{i=1,\dots,K},$ \\ \quad  \quad \quad \quad \quad \quad \quad \quad \quad 
                    \quad \quad \quad $\left(t_i^{j+1},q_i^{j+1} \right)_{i=1\dots,K} \Bigg)$
                \ENDFOR
            \UNTIL{convergence}
            \caption{Discrete $n$-geodesic in $\mathcal{B}$.
            Solving spline regression by minimizing (\ref{eq:regression_objective}) is denoted by \texttt{reg}.
            }
            \label{algo:discgeodesic}
        \end{algorithmic}
    \end{algorithm}
    
    Next, we discuss the computation of the \emph{discrete $n$-mean} of $S$ curves $B_1,\dots,B_S \in \mathcal{B}$, with which we approximate the common mean (curve). It is the spline $\overline{B} \in \mathcal{B}$ minimizing
    $G_n(p_0,\dots,p_K) := \sum_{s=1}^S E_n\left( (p_0^j,\dots,p_K^j)_{j=0,\dots,n}^{(s)} \right)$,
    s.t. $(p_0^n,\dots,p_K^n)^{(s)} = (p_0,\dots,p_K)^{(s)}$, $s=1,\dots,S$,
    where $(p_0^j,\dots,p_K^j)_{j=0,\dots,n}^{(s)}$ denotes the control points of the discrete geodesic between $B_s$ and $B(\cdot;p_0,\dots,p_K)$.
    It can be computed with an alternating optimization scheme. As an initialization of the control points of $\overline{B}$, we choose the means of the corresponding control points of the data curves. Then, in alternating fashion, discrete geodesics towards the mean are computed and, subsequently, the mean is updated by spline regression, since it has to lie ``in the middle'' of the innermost elements of the discrete geodesics. The procedure is summarized in Alg.~\ref{alg:mean}.

    \begin{algorithm}
        \hspace*{\algorithmicindent} \textbf{Input:} $B_1, \dots B_S \in \mathcal{B}$ with control points $(p_0^{(s)},\dots,p_K^{(s)})_{s=1,\dots,S}$,  discretization parameter $n \in \mathbb{N}$ \\
        \hspace*{\algorithmicindent} \textbf{Output:} Control points $(\overline{p}_0,\dots,\overline{p}_K)$ of the mean curve
        \begin{algorithmic}[H]
            \STATE $( \overline{p}_0,\dots,\overline{p}_K) \gets \Big(\texttt{mean}\big( p_0^{(1)},\dots,p_0^{(S)} \big), \dots,$ \\ 
            $\quad \quad \quad \quad \quad \quad \quad \quad \quad \quad \quad \quad \quad \quad \texttt{mean} \big( p_K^{(1)},\dots,p_K^{(S)} \big) \Big)$
            \REPEAT
                \FOR{$s=1,\dots,S$}
                    \STATE $(p_0^{j},\dots,p_K^{j})^{(s)}_{j=0,\dots,n} \gets$ \\
                    $\quad \quad \quad \quad \quad \quad \quad \quad \ \ \texttt{$n$-geo}( B(\cdot; \overline{p}_0,\dots,\overline{p}_K), B_s)$
                \ENDFOR
                \FOR{$s = 1, \dots, S$}
                    \FOR{$i=0,\dots,K$}
                        \STATE $\left( t_i^{(s)},q_i^{(s)} \right) \gets \left( \frac{iL}{K},B\big( \frac{iL}{K};p^{(s)}_0,\dots,p^{(s)}_K \big) \right)$
                    \ENDFOR
                \ENDFOR
                \STATE $( \overline{p}_0,\dots,\overline{p}_K) \gets \texttt{reg} \left( \left(t_i^{(s)},q_i^{(s)} \right)^{s=1,\dots,S}_{i=1,\dots,K} \right)$
            \UNTIL{convergence}
            \caption{Mean trajectory.
            Computation of Fr\'echet mean and $n$-geodesic are denoted \texttt{mean} and \texttt{$n$-geo}, respectively.
            }
            \label{alg:mean}
        \end{algorithmic}
    \end{algorithm}

\section{Experimental evaluation}

   \begin{figure*}
     \begin{center}
        \includegraphics[trim=.3cm .2cm 0cm .9cm, clip,width=0.79\textwidth]{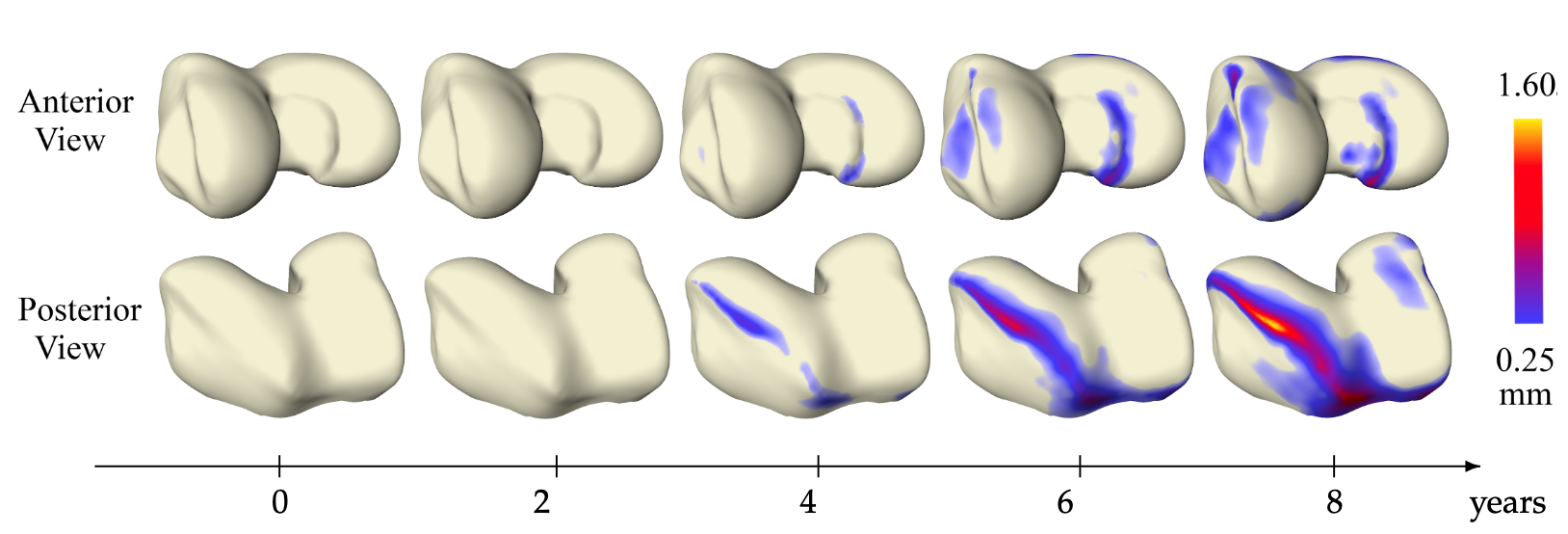}
        \caption{Mean of cubic femoral trends of 22 subjects evaluated at 5 equidistant points. The surface distance to the baseline (value of the computed mean at $t=0$) is color coded wherever the distance is larger than 0.25 mm.}
        \vspace{-1.5em}
     \label{fig:femur_cubic}
    \end{center}
    \end{figure*}

    Hierarchical models provide a principled way of analyzing longitudinal data. To demonstrate this, we perform group-wise analysis of femoral shape trajectories that have been collected from the Osteoarthritis Initiative (OAI).
    Furthermore, in order to demonstrate that the presented method is not limited to the estimation of average, group-level trends, we derive a statistical descriptor for shape trajectories in terms of the principal component scores (i.e., the coefficients encoding the trajectories within the basis of principal modes) and use it for \emph{trajectory} classification.

    The OAI is a longitudinal study of knee OA comprising (among others) clinical evaluation data and radiological images from 4,796 men and women of age 45-79 publicly available at \url{https://nda.nih.gov/oai/}.
    We determined three groups of shapes trajectories: HH (healthy, i.e.\ no OA), HD (healthy to diseased, i.e., onset and progression to severe OA), and DD (diseased, i.e., OA at baseline) according to the Kellgren--Lawrence score~\cite{KellgrenLawrence1957} of grade 0 for all visits, an increase of at least 3 grades over the course of the study, and grade 3 or 4 for all visits, respectively.
    Using an automatic segmentation approach~\cite{Ambellan_ea2019}, we extracted surfaces of the distal femora from the respective 3D weDESS MR images (0.37$\times$0.37 mm matrix, 0.7 mm slice thickness).
    For each group, we assembled 22 trajectories (all available data for group DD except one subject that exhibited inconsistencies, and the same number for groups HD and HH, randomly selected), each of which comprises shapes of all acquired MR images, i.e., at baseline, the 1-, 2-, 3-, 4-, 6-, and 8-year visits.
    As notion of shape space we employ the differential coordinates model~\cite{vonTycowicz_ea2018} that allows for closed-form evaluation of Riemannian operations and therefore facilitates fast and numerically robust processing.
    
    As a first application, we estimate a hierarchical model for the HD group. We employ cubic B\'ezier curves to model the individual trends. This choice is motivated by the findings in~\cite{Hanik_ea2020}, where cubic models were found to adequately capture the inherent nonlinear shape developments due to OA in a cross-sectional regression-based analysis.
    Time discrete computations are performed based on 2-geodesics---employing finer discretizations have been found to provide no further improvements for the dataset under study. The estimated group-level trend is visualized in Fig.~\ref{fig:femur_cubic}. The determined shape changes consistently expose OA related malformations of the femur, most prominently changes along the ridge of the cartilage plate that are characteristic regions for osteophytic growth. Note, that only minute bone remodeling can be observed for the first half of the captured interval, whereas bone malformations develop more rapidly after four years time. This behavior suggests that there are nontrivial higher order phenomena involved, which geodesic models cannot adequately describe.
    
    For the classification we train a simple support vector machine (linear kernel) on 65-dimensional descriptors (coefficients w.r.t.\ the PGA modes 
    from an approximated Gram matrix~\cite{Heeren_ea2018}) in a leave-one-out cross-validation setup.
    \begin{wrapfigure}{r}{0.22\textwidth}
        \footnotesize
        \begin{center}
            \vspace{-2.5em}
            \begin{tabular}{c | c c c}
                $\!\!\!\!\text{\footnotesize actual}$\textbackslash$^\text{\footnotesize pred.}\!\!\!$
                 & HH & DD & HD \\
                \hline
                HH & 19 & 1 & 2 \\ 
                DD & 2 & 11 & 9\\ 
                HD & 4 & 6 & 12 \\ 
            \end{tabular}
            \vspace{-2.5em}
        \end{center}
    \end{wrapfigure}
     The percentage of correctly classified trajectories is 64\%. The corresponding confusion matrix is given as inset. Performing the same experiment with a Euclidean model~\cite{Cootes_ea1995} results in 59\% correct classifications demonstrating the advantage of our Riemannian model.

    
\section{Conclusion and Future Work}

We presented a hierarchical statistical model that is based on intrinsic, higher-order B\'ezier splines and thus allows analyzing a wide range of phenomena with non-monotonous shape changes.
To the best of our knowledge this is the first Riemannian model that is neither bound to constraints of geodesicity nor based on non-parametric designs.

A promising direction for future work is to extend the hierarchical model presented to account for subject-specific shifts in the stage of evolution.
Adapting the proposed distance to partial trajectories could improve the estimation of group-level trends from longitudinal observations with highly varying inter-individual age range or disease stage coverage.

Furthermore, hierarchical models provide a principled way of analyzing longitudinal data.
In this regard, the presented method is not limited to estimating the average trends at, group-level, but can also be employed to assess the variance and principal modes of the distribution of the studied trajectories.
This opens up a multitude of applications such as hypothesis testing and Bayesian reconstruction, which we will address in the future.


\section{Compliance with Ethical Standards}
{ This research study was conducted retrospectively using human subject data made available in open access. Ethical approval was *not* required as confirmed by the license attached with the open access data.}

\section{Acknowledgments}
\vspace{-0.5em}
We are grateful for the funding by DFG\footnote{Deutsche Forschungsgemeinschaft (DFG) through Germany’s Excellence Strategy – The Berlin Mathematics Research Center MATH+ (EXC-2046/1, project ID: 390685689)} and BMBF\footnote{Bundesministerium f\"ur Bildung und Forschung (BMBF) through BIFOLD - The Berlin Institute for the Foundations of Learning and Data (ref. 01IS18025A and ref 01IS18037A)} as well as for the provision of the data set by the OAI\footnote{OAI is a public-private partnership comprised of five contracts (N01-AR-2-2258; N01-AR-2-2259; N01-AR-2-2260; N01-AR-2-2261; N01-AR-2-2262) funded by the National Institutes of Health, a branch of the Department of Health and Human Services, and conducted by the OAI Study Investigators. Private funding partners include Merck Research Laboratories; Novartis Pharmaceuticals Corporation, GlaxoSmithKline; and Pfizer, Inc. Private sector funding for the OAI is managed by the Foundation for the National Institutes of Health. This manuscript was prepared using an OAI public use data set and does not necessarily reflect the opinions or views of the OAI investigators, the NIH, or the private funding partners.}.
\vspace{-0.5em}


\bibliographystyle{IEEEbib}
\bibliography{bibliography}

\end{document}